\documentclass{revtex4}
\usepackage{epsfig}
\usepackage{graphicx}
\begin{document}
\bibliographystyle{revtex}
\preprint{FSU-HEP-2001-10-18}
\title{Associated $t\bar t{\cal H}$ production at a VLHC:\\
 measuring the top-quark
  Yukawa coupling}
\author{A.~Belyaev}
\email[]{belyaev@hep.fsu.edu}
\affiliation{Physics Department, Florida State University, Tallahassee,
FL 32306-4350}
\author{F.~Maltoni}
\email[]{maltoni@uiuc.edu}
\affiliation{Department of Physics, University of Illinois at
Urbana-Champaign, Urbana, IL 61801}
\author{L.~Reina}
\email[]{reina@hep.fsu.edu}
\affiliation{Physics Department, Florida State University, Tallahassee,
FL 32306-4350}
\date{\today}
\begin{abstract}
  Future hadron colliders will have the potential to measure some of
  the most relevant Higgs boson couplings with high precision. In this
  paper we investigate the potential of a Very Large Hadron Collider
  (VLHC) to measure the top-quark Yukawa coupling.
\end{abstract}
\maketitle
\section{Introduction}
\label{sec:intro}
As part of the Snowmass effort to investigate the physics potential of
future hadron colliders, we have addressed the problem of how some of
the most relevant precision measurements of Higgs boson physics could
benefit from the very high energy and statistics of these future
facilities.

We imagine a scenario in which one or more Higgs bosons have been discovered at
either the Fermilab Tevatron or the CERN Large Hadron Collider (LHC), and a
rich program of Higgs boson physics has already been developed. We then work
under the assumption that precise determinations of the Higgs boson mass(es)
and width(s), as well as determinations of various Higgs boson production cross
sections, branching ratios, and ratios of Higgs boson couplings within a
10-20\% uncertainty are available.  The next generation of colliders will then
play a crucial role in getting to a more precise determination of the Higgs
boson couplings, therefore constraining its nature. It has been shown that an
$e^+e^-$ Linear Collider, operating with high luminosity, can reach precisions
of a few percents on all Higgs boson couplings except the Higgs boson
self-couplings~\cite{Abe:2001wn,TDRTESLA:2001}.  The question is therefore what
is the corresponding potential of a next generation hadron collider like a Very
Large Hadron Collider (VLHC).

Among the most important Higgs boson couplings, the Higgs-boson coupling to
the top quark plays a special role. Because of the intriguingly large size of
the top-quark mass, this coupling is largely enhanced with respect to all other
Yukawa couplings and could shed some light on the obscure pattern of fermion
mass generation and electroweak symmetry breaking.

In this context, it is interesting to assess the precision with which
the top-quark Yukawa coupling could be measured at a $pp$ VLHC,
running at center of mass energies of $\sqrt{s}\!=\!40,100,200$~TeV
respectively.  The golden mode for this measurement is the associated
production of a Higgs boson with a pair of top-antitop quarks,
$pp\rightarrow t\bar t {\cal H}$,~\cite{Marciano:1991qq}.  The Higgs
boson is radiated either from the top or from the antitop quarks and
the cross section is directly proportional to the top-quark Yukawa
coupling~\cite{Ng:1984jm,Kunszt:1984ri}.  We mainly focus on a
Standard Model (SM) like Higgs boson (${\cal H}\!=\!h_{SM}$), giving
only some qualitative indication of how the analysis could be
generalized to the Minimal Supersymmetric Standard Model (MSSM) Higgs
sector (${\cal H}\!=\!h^0,H^0,A^0$).  In our analysis we consider the
Higgs boson decaying into $b\bar b$, $\gamma\gamma$, and
$\tau^+\tau^-$ and we determine the significance of the signal over
the background in these three cases.  As a result, all three channels
turn out to be viable, even for fairly low integrated luminosities,
providing a determination of the top-quark Yukawa coupling at the few
percent level over a large range of Higgs boson masses.

The layout of our presentation is as follows. The characteristics of
the the associated production of a SM like Higgs boson in
$pp\rightarrow t\bar t {\cal H}$ are described in
Sec.~\ref{sec:signal}. In Sec.~\ref{sec:sig_vs_back} we compare signal
and background, in the SM, for the three Higgs boson decay channels
discussed above, and estimate the relative error with which the SM
top-quark Yukawa coupling could be measured at a VLHC. We also give
some qualitative indications of how the results could change if the
MSSM Higgs sector is considered. Sec.~\ref{sec:concl} contains our
conclusions.
\begin{figure}[htb]
  \begin{minipage}[b]{.46\linewidth}
    \hspace{-1.truecm}
    \includegraphics[scale=.6]{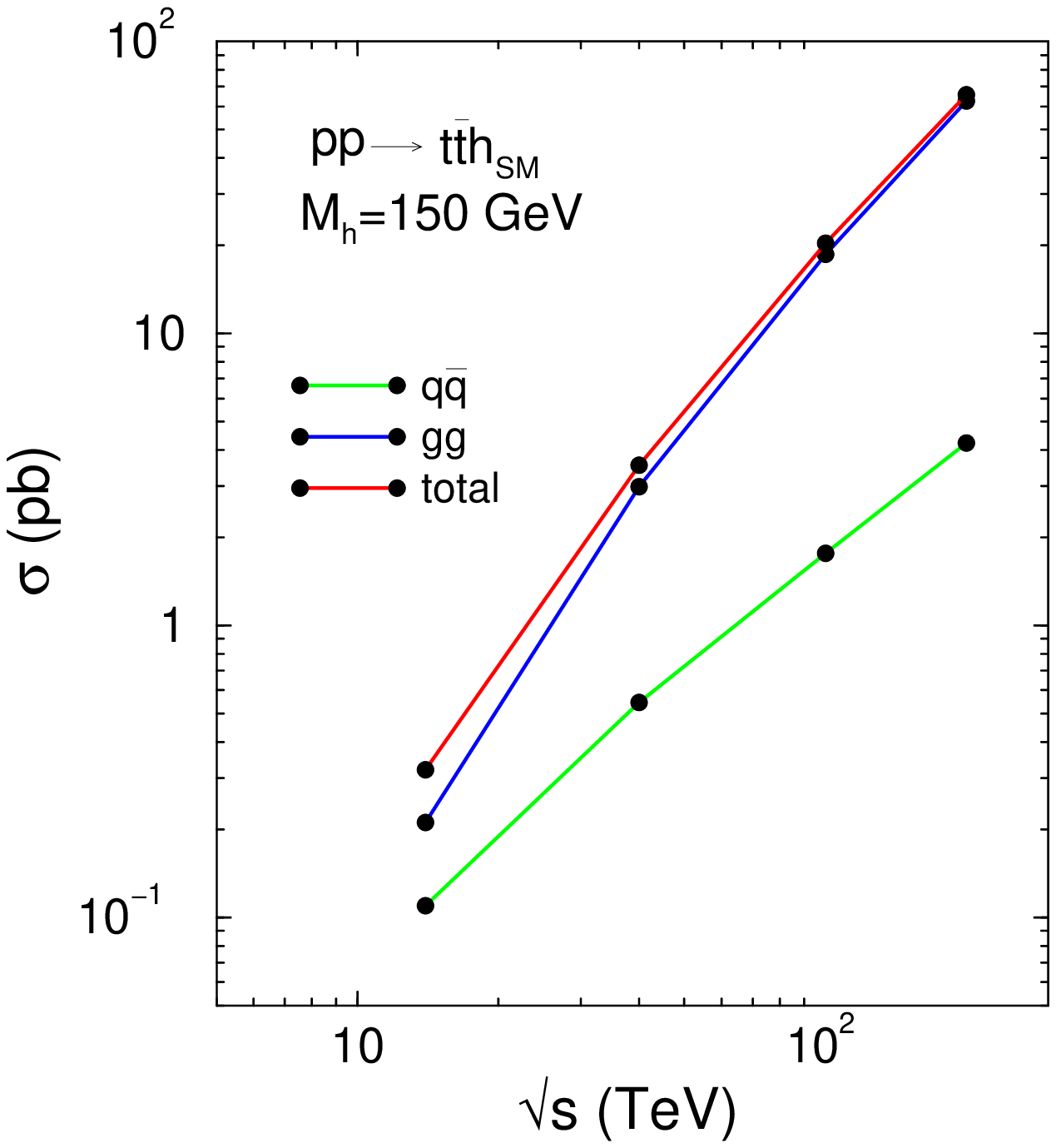}
    \caption{Relative contributions of the
      $q\bar q$ and $gg$ parton level channels to the total cross
      section for $pp\rightarrow t\bar th_{SM}$ as functions of the
      center of mass energy $\sqrt{s}$.}
    \label{fig:sigma_qq_gg}
  \end{minipage}\hspace{0.5truecm}
  \begin{minipage}[b]{.46\linewidth}
    \hspace{-0.5truecm}
    \includegraphics[scale=.6]{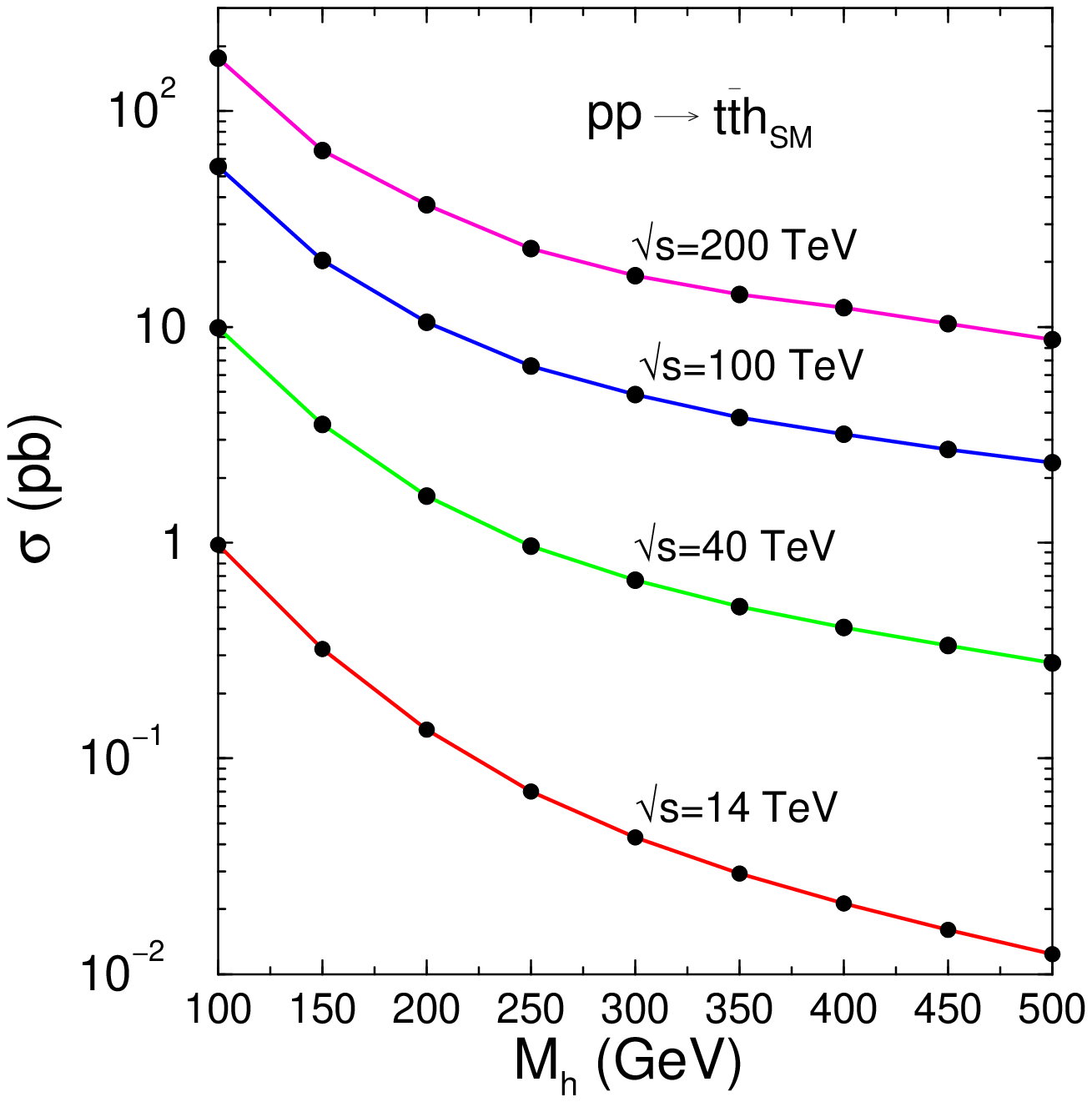}
    \caption{Total cross section for $pp\rightarrow t\bar t h_{SM}$
      as a function of $M_{h_{SM}}$, for various center of mass
      energies.}
    \label{fig:sigmatot_s_mh}
    \vspace{0.35truecm}
  \end{minipage}
\end{figure}

\section{Signal}
\label{sec:signal}
The total hadronic cross section for $pp\rightarrow t\bar t {\cal H}$
(${\cal H}\!=\!h_{SM},h^0,H^0,A^0$) consists of two parton level
sub-processes: $q\bar q\rightarrow t\bar t {\cal H}$ and
$gg\rightarrow t\bar t {\cal H}$.  Taking ${\cal H}\!=\!h_{SM}$ for
illustrative purposes, we plot in Fig.~\ref{fig:sigma_qq_gg} the
relative contribution of the two subprocesses, for $M_{\cal
  H}\!=\!150$ GeV. As expected, the $gg$ contribution dominates as the
center of mass energy is increased.

In Fig.~\ref{fig:sigmatot_s_mh} we also show the dependence of the
total cross section from $M_{\cal H}$, again when ${\cal
  H}\!=\!h_{SM}$, for $\sqrt{s}\!=\!14$~TeV (LHC) and
$\sqrt{s}\!=\!40,100,200$~TeV (VLHC).  For the highest center of mass
energy considered in this paper, $\sqrt{s}\!=\!200$ TeV, the total
cross section is enhanced by two to three orders of magnitude with
respect to the corresponding cross section at the LHC, depending on
the Higgs boson mass.

All the results presented in this paper are obtained using tree-level
cross sections, both for the signal and for the backgrounds,
calculated using CTEQ4L~\cite{Lai:1995bb} parton distribution
functions and the strong coupling constant $\alpha_s(\mu)$ at
one-loop.  As usual, tree-level cross sections have a very large
renormalization/factorization scale dependence and, as a result, a
large uncertainty. At present, only the next-to-leading QCD
corrections to the signal are
known~\cite{Beenakker:2001rj,Reina:2001sf,Reina:2001bc}.  Since we do
not aim at a precise determination of the cross section, but at a
study of signal vs. background, we prefer to consistently use only
quantities calculated at leading order, without including any
K-factors.  The renormalization and factorization scales have been set
to a common value $\mu\!=\!m_t+M_{\cal H}/2$, with $m_t\!=\!174$~GeV.

\section{Signal vs. Background for various decay channels}
\label{sec:sig_vs_back}

\begin{figure}[htb]
\noindent
\mbox{         \epsfig{file=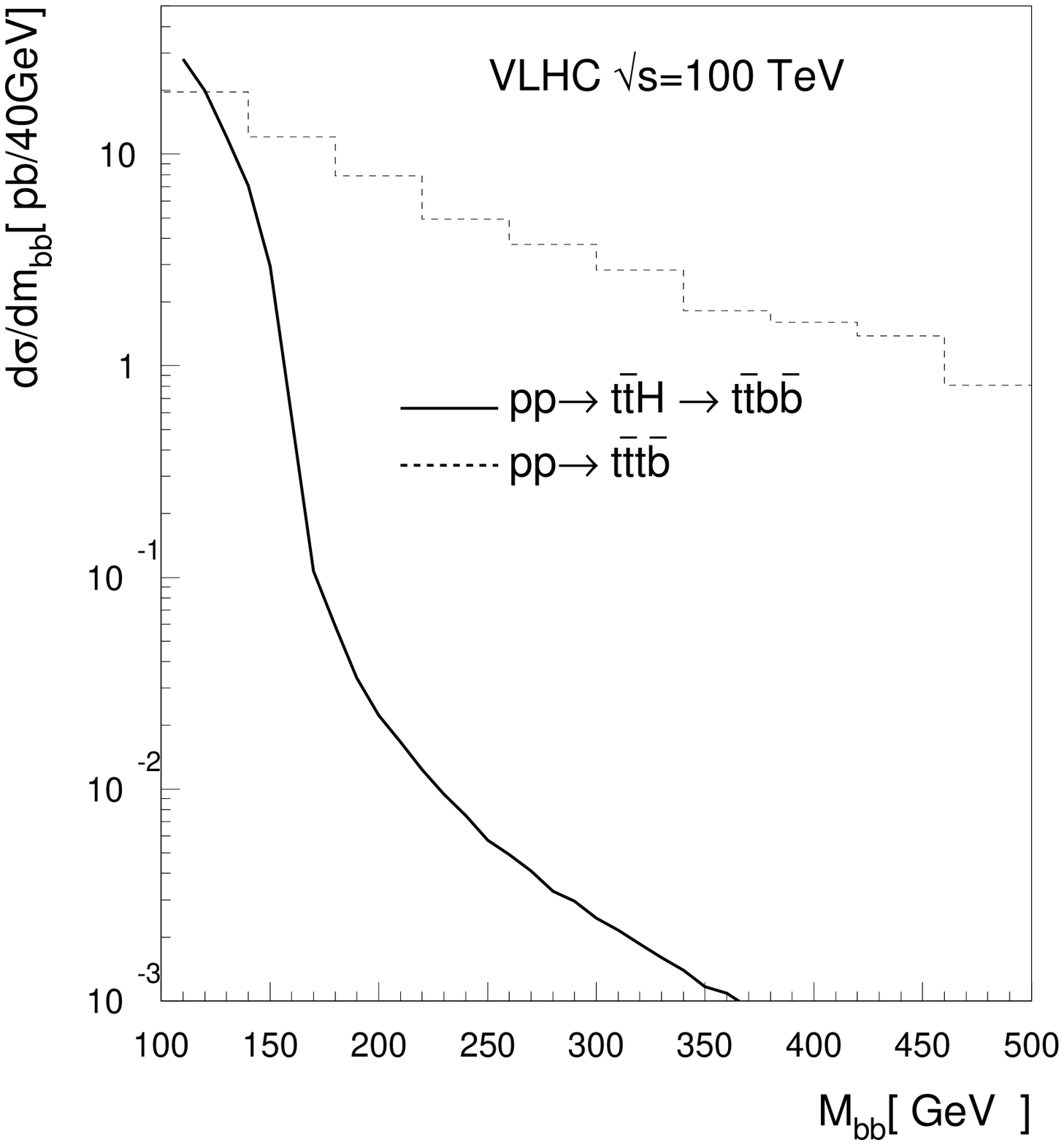,width=0.35\textwidth}
 \hspace*{-0.5cm}\epsfig{file=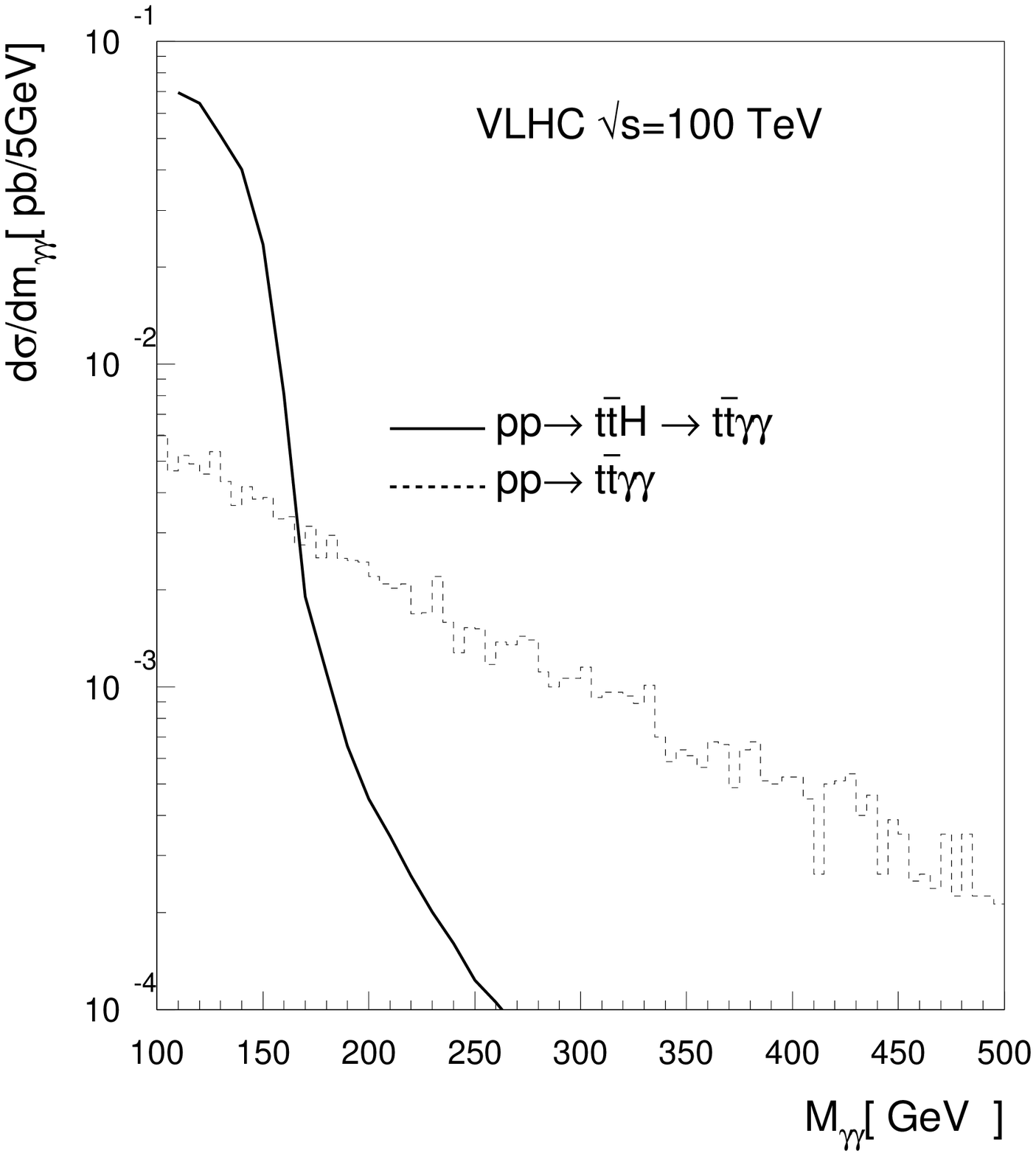,width=0.35\textwidth}
 \hspace*{-0.5cm}\epsfig{file=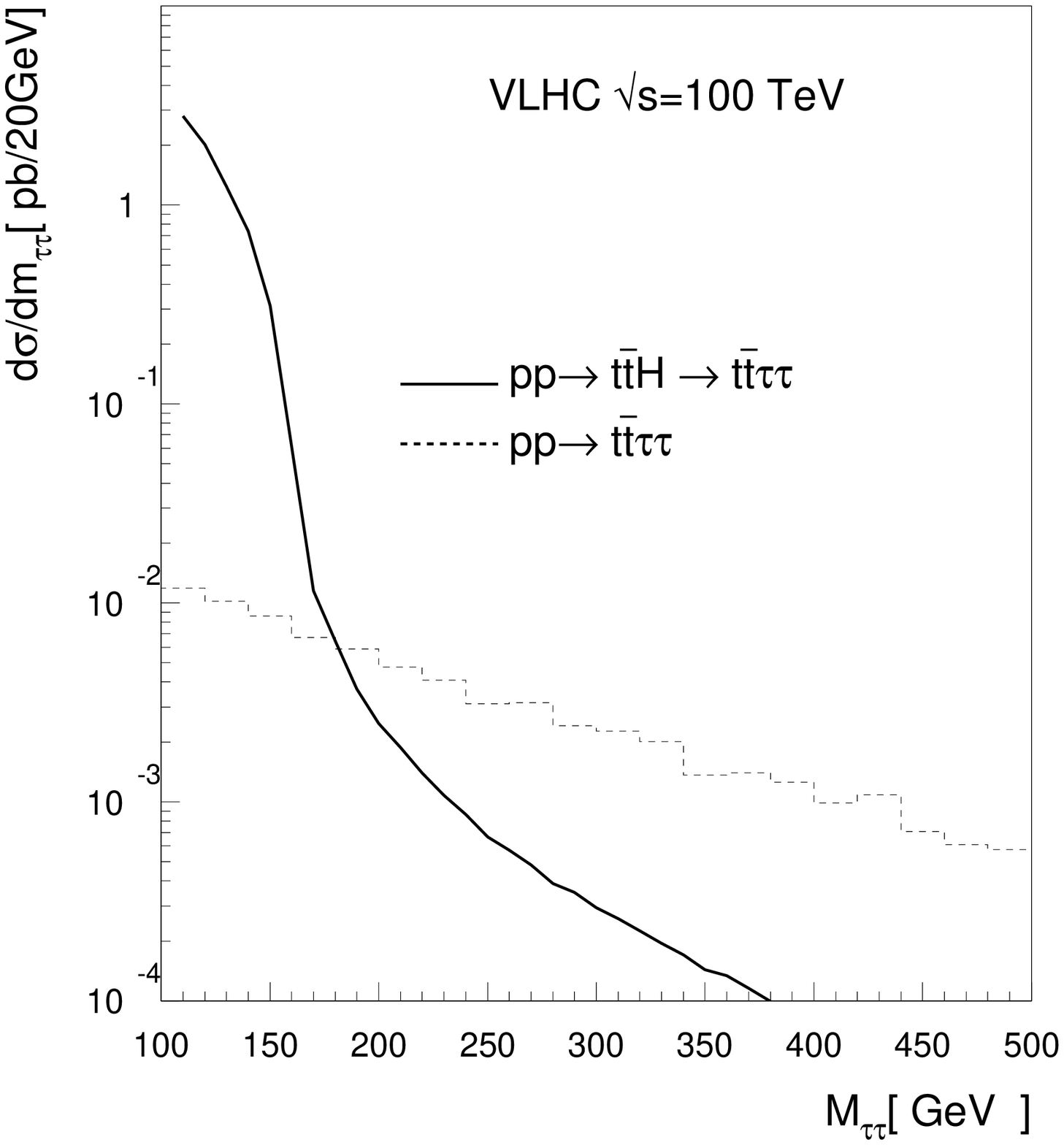,width=0.35\textwidth}}
\caption{Cross sections for both signal (solid line) and irreducible background
  (dashed histogram) for the three signatures $t\bar tXX$ for
  $X=b,\gamma,\tau$, as functions of the invariant mass $M_{XX}$ of
  the $XX$ pair, at a VLHC with $\sqrt{s}=100$ TeV. The background
  cross sections are plotted in $M_{XX}$ bins of $40$, $5$, and $20$~GeV,
  respectively. To simulate the detector acceptance we require
  $p_T>25$ GeV and $|\eta|<3$.
  \label{fig:signalvsbkgs} }
\vspace*{-0.5cm}
\end{figure}

In this section we present some studies of the irreducible backgrounds and
discuss the expected precision with which a measurement of the SM top-quark
Yukawa coupling can be performed at a VLHC. For illustration purposes, we
consider the case of a VLHC operating at $\sqrt{s}\!=\!100$~TeV. In
Fig.~\ref{fig:signalvsbkgs} we compare the cross sections for the signal,
$pp\rightarrow t\bar t h_{SM}$, and for the irreducible backgrounds consisting
of $t\bar{t}XX$ production with $X=b,\gamma, \tau$~\footnote{Even though we do
not include in this study other decay modes, such as, for instance, $h_{SM} \to
WW,ZZ$, they are potentially interesting and more detailed analysis are in
progress.}.  The cross sections for the signal include the branching ratios
$h_{SM}\to XX$, calculated using HDECAY~\cite{Djouadi:1998yw}. In order to take
into account finite mass resolution effects, the background events are plotted
in bins of $40$, $5$, and $20$~GeV for $b$, $\gamma$, and $\tau$ respectively.
To simulate the detector acceptance, the decay products of the Higgs boson are
required to have a transverse momentum $p_T>25$ GeV and a pseudorapidity
$|\eta|<3$. The set of parton distribution functions is CTEQ4L and the
renormalization and factorization scales are set equal to $m_t+M_{XX}/2$, where
$M_{XX}$ is the invariant mass of the $XX$ pair.
Qualitatively, the signal to background ratios are similar to those
expected at the LHC~\cite{Beneke:2000hk}. For the leading decay
channel $h_{SM}\rightarrow b\bar b$ the QCD background is comparable
to the signal, while for the $h_{SM}\rightarrow \gamma \gamma$ and
$h_{SM}\rightarrow\tau^+\tau^-$ decay channels the background is
small, if not negligible. However, for these last two channels, the
advantage of a VLHC over the LHC is manifest. Already with
100~fb$^{-1}$ of integrated luminosity, at a VLHC with
$\sqrt{s}\!=\!100$~TeV, the number of signal events is increased by
about a factor of $50-100$ (for $M_{h_{SM}}\!=\!100-200$~GeV) with
respect to the LHC, therefore allowing studies that are statistically
limited at the LHC. Even for the $h_{SM}\to b\bar{b}$ decay channel,
assuming that efficiencies similar to those at the LHC could be
attained, the significance of the signal (directly related to the
accuracy with which the top Yukawa coupling can be measured) would be
increased by a factor $\simeq \sqrt{50}$.
\begin{table*}[hbt]
\caption{ Number of signal events for $t\bar{t}h_{SM}$,
  $M_{h_{SM}}=130$~GeV, and irreducible $t \bar{t} XX$ background
  events with $X=b,\gamma,\tau$, at a VLHC with $\sqrt{s}=100$ TeV and
  integrated luminosity of 100 fb$^{-1}$. Same conventions as in
 Fig.~\ref{fig:signalvsbkgs}. In the total number of events the
 branching ratios for $h_{SM}\to XX$ with $X=b,\gamma,\tau$ are included.
 In the second line the top-quark and $\tau$ branching ratio into final
 states are included. Detector and reconstruction efficiencies are estimated
 from LHC studies (e.g.,~\cite{Beneke:2000hk}) and inserted in the third line.
 In the last line, (one half of) the relative error on the measurement of the cross section
 is quoted. This is directly related to the precision on the extraction of the Yukawa coupling of the top,
 $\delta y_t/y_t=\sqrt{(\delta \sigma/2\sigma)^2+ (\delta y_X/y_X)^2 }$, and
 corresponds to it when the uncertainty on the branching
 ratios of the Higgs into the final states is negligible.}
\label{yukawa}
\medskip
\addtolength{\arraycolsep}{1cm}
\renewcommand{\arraystretch}{1.8}
\begin{center}
\begin{tabular}{|c|cc|cc|cc|}
\hline\hline
& \multicolumn{2}{c|}{$b \bar{b}$} &
\multicolumn{2}{c|}{$\gamma\gamma$}&
\multicolumn{2}{c|}{$\tau^+  \tau^- $} \\
& S & B & S & B & S & B\\
\hline
\# events &
$1.2\cdot10^6$ & $2.4\cdot10^6$ & $3.8\cdot10^3$ & $5.7\cdot10^2$ &
$9.0\cdot 10^4$ & $1.2\cdot10^3$\\
w/ BR's &
$1.8\cdot10^5 $ & $3.6\cdot10^5$ & $5.7\cdot10^2$ & $ 86 $ &
$5.5 \cdot 10^3$ & 74 \\
w/ efficiencies  &
$3.4\cdot10^3$ & $6.7\cdot10^3$ & $60$ &  $9$  & $ 2.1\cdot10^2 $ & 3 \\
\hline
S/B & \multicolumn{2}{c|}{0.5} & \multicolumn{2}{c|}{6} &
\multicolumn{2}{c|}{$70   $}\\
${\delta\sigma}/2\sigma$ (\%) & \multicolumn{2}{c|}{1.5} & \multicolumn{2}{c|}{7} &
\multicolumn{2}{c|}{3.5}\\
\hline\hline
\end{tabular}
\end{center}
\vspace*{-0.2cm}
\end{table*}
A first estimate of the precision with which the SM top-quark Yukawa
coupling could be measured at a VLHC is given in Table~\ref{yukawa},
for $\sqrt{s}\!=\!100$~TeV and $M_{h_{SM}}\!=\!130$ GeV.  In our
analysis we assume 100~fb$^{-1}$ of total integrated luminosity,
corresponding to roughly one year of running of a VLHC with luminosity
${\cal L}\!=\!10^{34}\,\mbox{cm}^{-2}\mbox{s}^{-1}$.  In analogy with
similar studies performed for the LHC ~\cite{TDRCMS:1994,
  TDRATLAS:1999, Beneke:2000hk}, we consider a sample where one top
quark decays leptonically, in order to have an unambiguous lepton tag,
and the other top decays hadronically. We apply a $t\bar t$ pair
reconstruction efficiency $\epsilon_{t\bar t}\!=\!0.15$.  We also use
a $b$ tagging efficiency $\epsilon_b\!=\!0.6$, a $\tau$ tagging
efficiency $\epsilon_\tau\!=\!0.6$, and a photon identification
efficiency $\epsilon_\gamma\!=\!0.9$.  Moreover, in order to account
for the efficiency of the invariant mass finite cut imposed by the
binning procedure, we multiply the results for the $t\bar tb\bar b$
and $t\bar t\tau^+\tau^-$ signatures by
$\epsilon_{mc}^{b,\tau}\!=\!0.7$, and the results for the $t\bar
t\gamma\gamma$ signature by $\epsilon_{mc}^{\gamma}\!=\!0.9$ .  The
$t\bar tb\bar b$ signature is further reduced by a factor
$\epsilon_{comb}=0.5$, to take into account the combinatorics due to
the four $b$ quarks in the final state.  Finally, we only consider the
$t\bar t\tau^+\tau^-$ signature where the $\tau$'s decay hadronically
and we therefore multiply by Br$(\tau\to\mbox{hadrons})\!=\!0.63$ for
each $\tau$ lepton in the final state~\footnote{In a more complete
  analysis other data samples should be added to this channel. For
  instance, the case where one $\tau$ decays leptonically, providing
  the lepton tag, and both the top quarks decay hadronically has a
  comparable rate.}.

As a result, the statistical error on the $t\bar th_{SM}$ production
cross section for a SM Higgs boson with mass 130 GeV is at the
percentage level for both the $t\bar tb\bar b$ and $t\bar
t\tau^+\tau^-$ signatures, while for the $t\bar t\gamma\gamma$
signature is around 10\%. The precision with which the top-quark
Yukawa coupling can be extracted from the measured $t\bar th_{SM}$
cross section depends on the accuracy on the $h_{SM}\rightarrow XX$
branching ratios (see Table~\ref{yukawa}). Assuming that all other
Higgs boson couplings entering our analysis have been determined with
very good precision, then all three signatures allow a measurement of
the top Yukawa coupling with precision better than 10\%, for Higgs
boson masses up to 150 GeV. We note that the level of precision
obtained is comparable with the precision that could be attained at a
high energy Linear Collider, running at the optimal center of mass
energy of $\sqrt{s}\!=\!800$~GeV, when $10^3$~fb$^{-1}$ of integrated
luminosity are used \cite{Baer:1999ge,Juste:1999af}. On the other
hand, our results are obtained using a quite low integrated
luminosity, $10^2$~fb$^{-1}$, and could therefore be further improved
by more available statistics.

\begin{figure}[tb] \noindent \mbox{
\epsfig{file=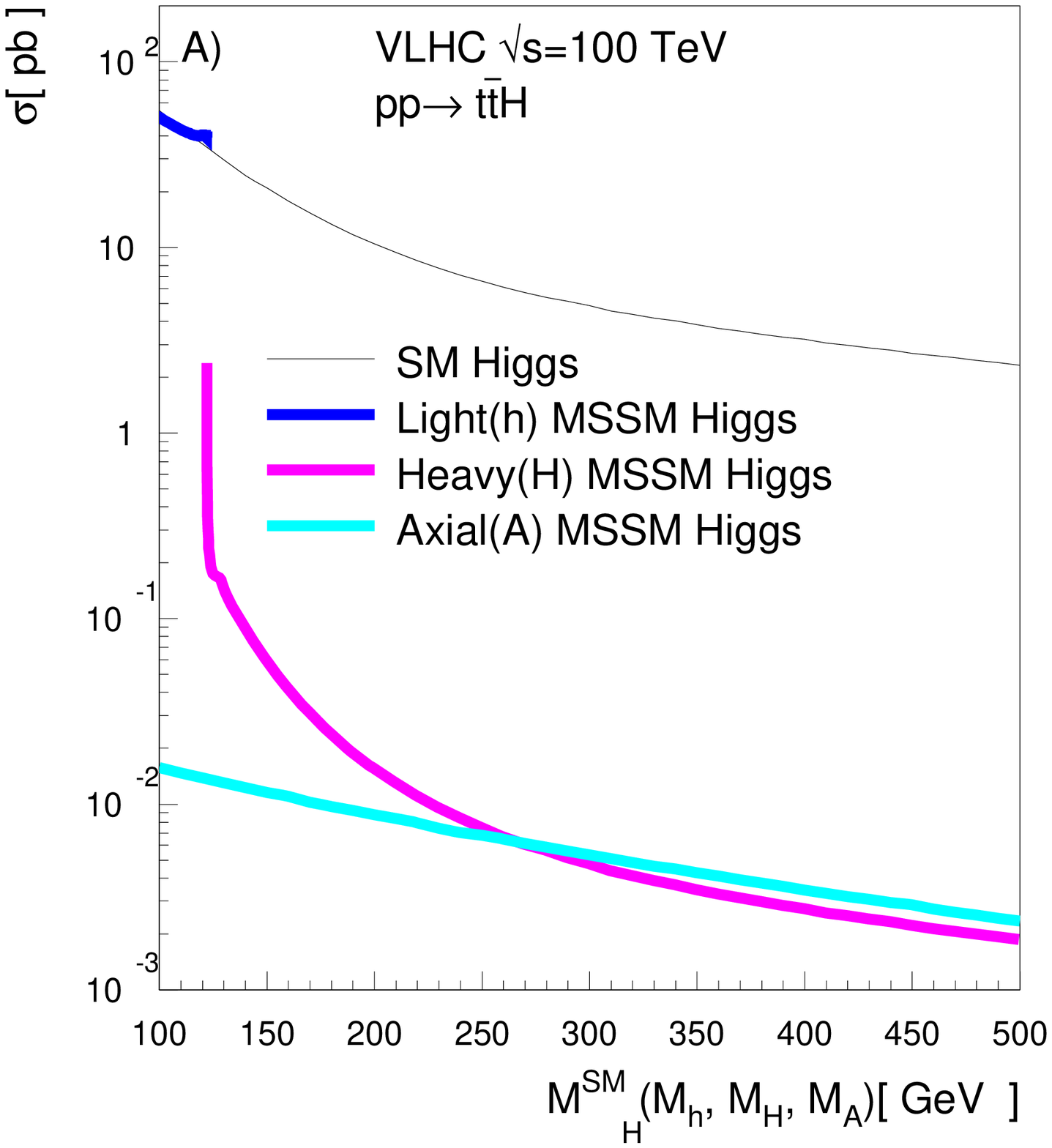,width=0.4\textwidth}
                    \epsfig{file=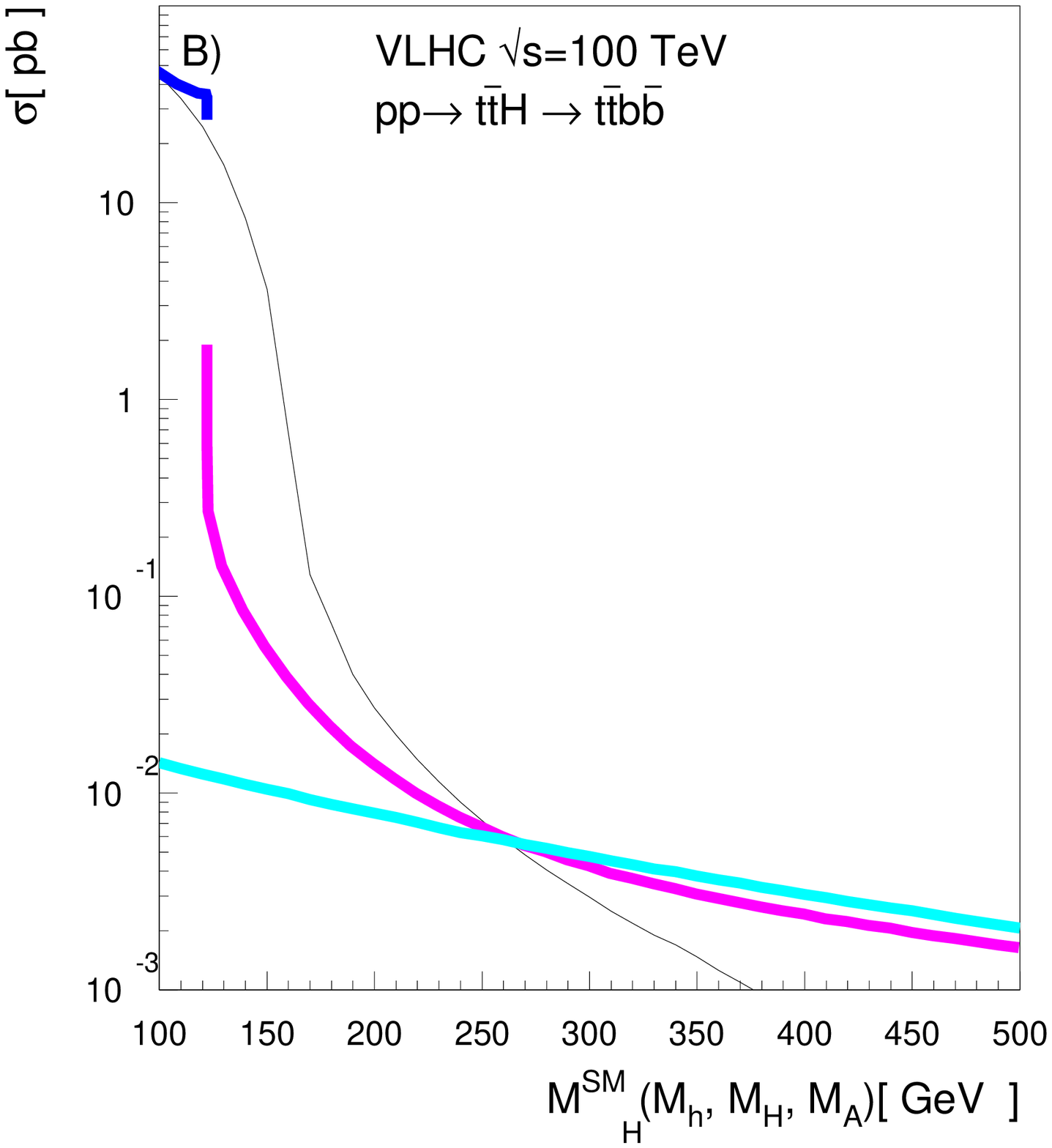,width=0.4\textwidth}}
\mbox{              \epsfig{file=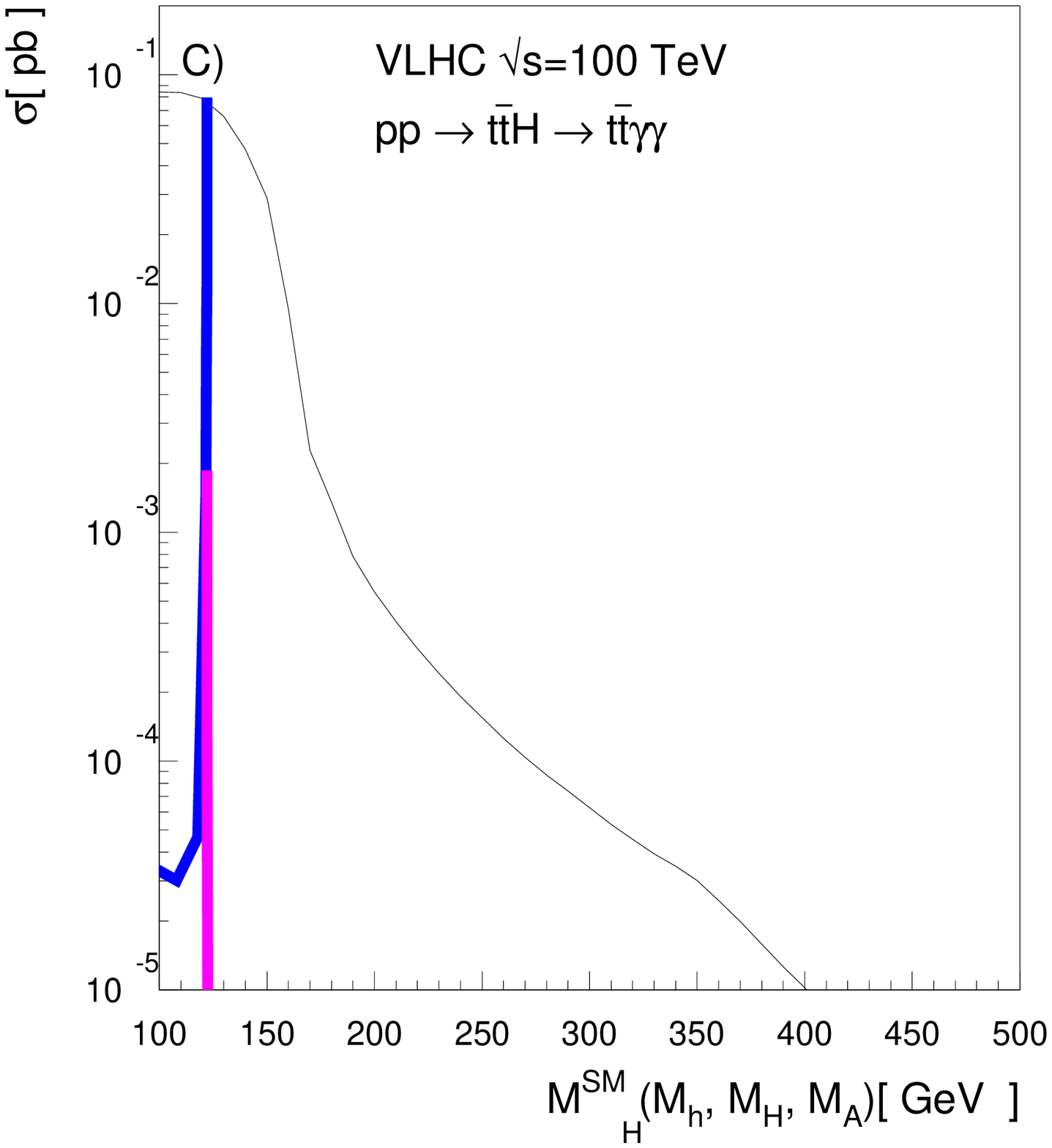,width=0.4\textwidth}
                    \epsfig{file=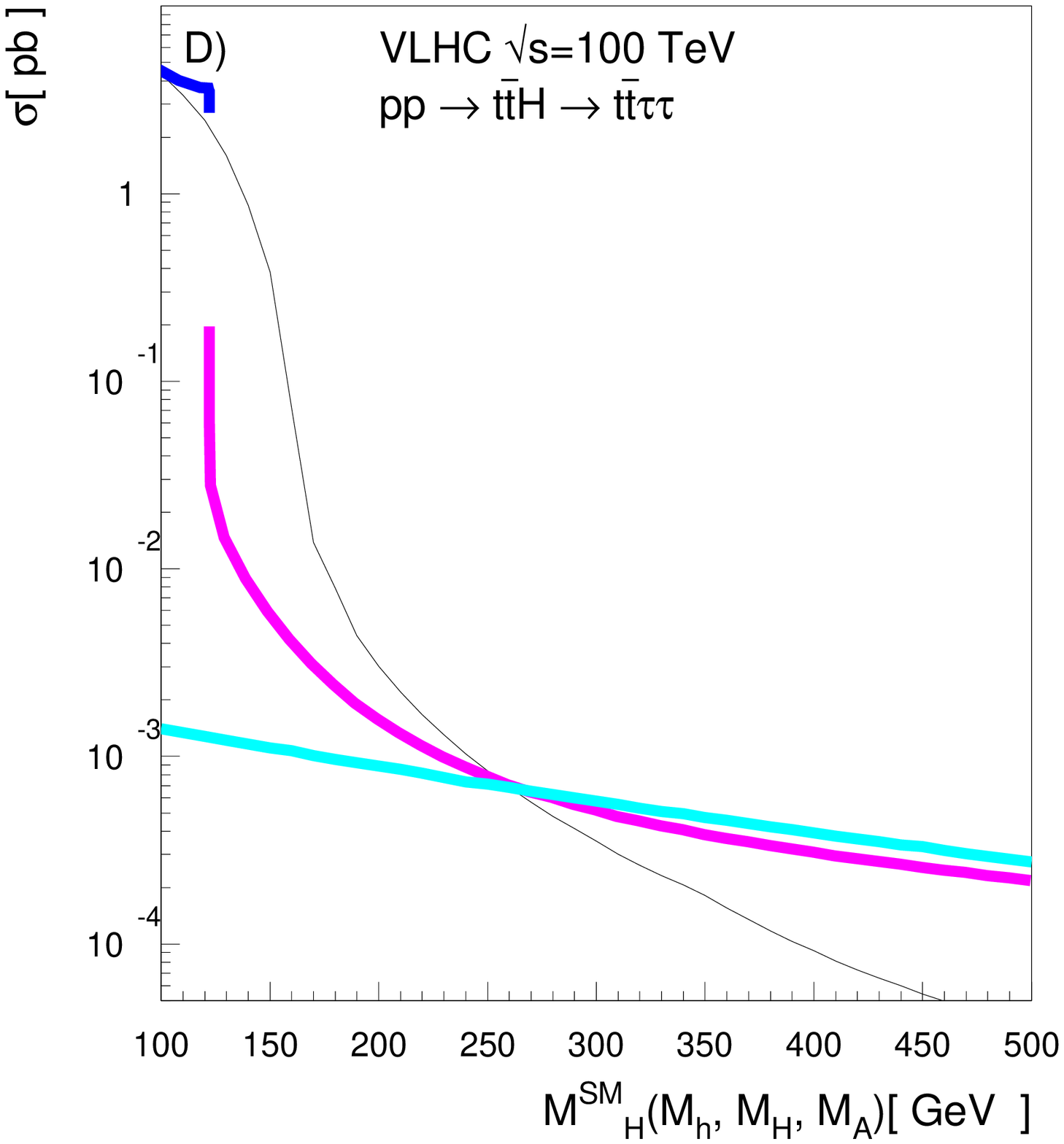,width=0.4\textwidth}}
\caption{\label{fig:mssm} Cross sections
  for $t\bar{t}h_{SM}$ (thin solid line) and $t\bar{t}{\cal
    H}$(${\cal H}=h^0,H^0$ and $A^0$) production (thick black,
  dark-grey and light-grey lines), at a VLHC with $\sqrt{s}=100$ TeV:
  A) total cross sections; B)-D) cross sections times branching ratios
  for ${\cal H}\rightarrow b\bar b,\gamma\gamma,\tau^+\tau^-$
  respectively.}
\end{figure}

Finally, it is worth having a quick look at $t\bar t{\cal H}$
production in the MSSM (for ${\cal H}\!=h^0,H^0,A^0$).  For the MSSM
scenario we assume some ``typical'' set of parameters: $\tan\beta=40$,
$m_{\tilde q}=500$~GeV, $\mu=300$~GeV, $A_t\!=\!A_b\!=\!A$ with
$A=\mu/\tan\beta +\sqrt{6}m_{\tilde q}$, according to the ``maximal
mixing" scenario where loop corrections maximize the light Higgs boson
mass.  The total cross sections for the signal $t\bar t {\cal H}$ as
well as the total cross sections multiplied by the branching ratios
for ${\cal H}\rightarrow b\bar b,\gamma\gamma,\tau^+\tau^-$ are
presented in Fig.~\ref{fig:mssm}.

The cross sections for $t\bar{t}h^0$ and $t\bar{t}h_{SM}$ are close to
each other in the narrow, but crucial, region up to $M_{\cal
  H}\!=\!120$~GeV or slightly above that. However, we note that, when
the branching ratios are included, both the $t\bar tb\bar b$ and $t\bar
t\tau^+\tau^-$ MSSM signatures for $h^0$ are 30-40\% higher than the
corresponding SM signatures. Above 120~GeV only the $t\bar tH^0$ and $t\bar
tA^0$ associated production can take place.  The $t\bar t H^0$ signal is
suppressed by a factor of $\cos\beta^2$ compared to the corresponding SM
signal. It is similar to the light Higgs boson ($h^0$) signal only in the small
region around 120~GeV, where all three MSSM Higgs bosons are degenerate in
mass. For $M_{\cal H}\!>\!120$~GeV the $t\bar tH^0$ cross section drops rapidly
and becomes comparable to the $t\bar tA^0$ cross section for Higgs boson masses
above 200~GeV.  The cross sections for both $t\bar tH^0$ and $t\bar tA^0$ above
120~GeV are 2-3 orders of magnitude below the SM cross section.  However, when
we take into account the corresponding Higgs boson decay branching ratios, the
situation can be very different. For instance, when ${\cal
H}\rightarrow\tau^+\tau^-$, the MSSM cross sections start dominate the SM one
over the entire mass region $M_{\cal H}>200$~GeV.  This happens because the
MSSM ${\cal H}\rightarrow\tau^+\tau^-$ branching ratio is significant over the
entire Higgs boson mass region for high $\tan\beta$. With this respect, the
$t\bar t\tau^+\tau^-$ supersymmetric signature (summed over the $t\bar tA^0$
and $t\bar tH^0$ channels) could be interesting in the high MSSM Higgs boson
mass range ($M_{\cal H}\!\ge\!200$~GeV). Since however, even at a VLHC, this
channel appears to be statistically limited, it will require a large integrated
luminosity. We note that the MSSM $t\bar tb\bar b$ signature also dominates
over the SM one for large Higgs masses, but in this case, contrary to $t\bar
t\tau^+\tau^-$, the background is overwhelming.

\section{Conclusions}
\label{sec:concl}
In this note we have studied the precision with which the top-quark
Yukawa coupling could be determined at a $pp$ VLHC through the
measurement of the cross section for the process $pp\rightarrow t\bar
t {\cal H}$, with the Higgs boson subsequently decaying into $b\bar
b$, $\gamma\gamma$, and $\tau^+\tau^-$. We have mainly focused on a SM
like Higgs boson, but have also looked at some interesting MSSM
signatures, for both light and heavy Higgs bosons.  Assuming that the
branching ratios of the Higgs into the final states were known with very good precision,
each of the three Higgs boson decay channels could provide a determination
of the top-quark Yukawa coupling at the few percent level, over a
large range of Higgs boson masses.  In particular, the $\gamma\gamma$
and $\tau^+\tau^-$ channels, which will be statistically limited at
the LHC, are at the VLHC very clean and already significant with just
100~fb$^{-1}$ of integrated luminosity. This could be extremely
useful, even under the pessimistic assumption that some of the Higgs
boson branching ratios had still to be determined by the VLHC era.
For instance, the determination of $y_t$ from the $\gamma\gamma$
and/or $\tau^+ \tau^-$ channels, could be used to extract the
branching ratio of the Higgs into $\bar{b} b$, which will be hard to
measure directly at the LHC. Or, one could directly check that the
ratio $\Gamma(H\rightarrow b\bar b)/\Gamma(H\rightarrow\tau^+\tau^-)$
behaves as $m_b^2/m_\tau^2$, and subsequently extract the top Yukawa
coupling $y_t$ by following a strategy similar to the one suggested in
Ref.~\cite{Zeppenfeld:2000td}.

\begin{acknowledgments}
  We are grateful to David Rainwater for his comments and suggestions.
  The work of A.B. and L.R. (F.M.) is supported in part by the U.S. Department
  of Energy under contract No.~DE-FG02-97ER41022 (DE-FG02-91ER40677).
\end{acknowledgments}
\bibliography{E4_reina_1010}
\end{document}